\documentclass[preprint]{revtex4}%
\usepackage{amssymb}
\usepackage{amsfonts}
\usepackage{amsmath}
\usepackage{graphicx}%
\providecommand{\U}[1]{\protect\rule{.1in}{.1in}}
\begin{document}
\title{Exact quasinormal modes for a special class of black holes}
\author{Julio Oliva$^{1}$, Ricardo Troncoso$^{2,3}$}
\affiliation{$^{1}$Instituto de F\'{\i}sica, Facultad de Ciencias, Universidad Austral de Chile.}
\affiliation{$^{2}$Centro de Estudios Cient\'{\i}ficos (CECS), Casilla 1469, Valdivia, Chile.}
\affiliation{$^{3}$Centro de Ingenier\'{\i}a de la Innovaci\'{o}n del CECS (CIN), Valdivia, Chile.}
\preprint{CECS-PHY-06/13}

\begin{abstract}
Analytic exact expressions for the quasinormal modes of scalar and
electromagnetic perturbations around a special class of black holes are found
in $d\geq3$ dimensions. It is shown that, the size of the black hole provides
a bound for the angular momentum of the perturbation. Quasinormal modes appear
when this bound is fulfilled, otherwise the excitations become purely damped.

\end{abstract}
\maketitle

\section{Introduction}

Small perturbations of the geometry or matter fields drained by a black hole
give rise to the so-called quasinormal modes. Since the spectrum is
independent of the initial conditions, and it is characterized only on the
parameters of the black hole and on the fundamental constants of the system,
it contains relevant information about the intrinsic properties of the black
hole. Furthermore, by virtue of the AdS/CFT correspondence, quasinormal modes
determine the relaxation time scale of the thermal states in the dual theory
\cite{Horowitz-Hubeny}. Exact results about quasinormal modes and frequencies
certainly help to have a deeper understanding of this phenomenon.
Nevertheless, only few analytic results are known \cite{Analytic1}%
-\cite{Analyticf}. Our purpose is to report that there is a special class of
black holes in $d\geq3$ dimensions, for which exact analytic expressions for
the quasinormal modes of scalar and electromagnetic perturbations can be
obtained. One can see that the size of the black hole provides a bound for the
angular momentum of the perturbation, such that quasinormal modes appear when
this bound is fulfilled; otherwise the excitations become purely damped.

Let us consider the following spacetime in $d$-dimensions%
\begin{equation}
ds^{2}=-\frac{1}{l^{2}}\left(  r^{2}-r_{+}^{2}\right)  dt^{2}+\frac
{l^{2}dr^{2}}{r^{2}-r_{+}^{2}}+r^{2}d\Sigma_{d-2}^{2}\ ,\label{bh}%
\end{equation}
where $l$ is the AdS radius and $d\Sigma_{d-2}^{2}$ stands for the metric of a
smooth "base" manifold $\Sigma_{d-2}$ of $d-2$ dimensions which can be assumed
to be compact and orientable. This metric describes an asymptotically locally
AdS black hole whose event horizon is located at $r=r_{+}$. In $d=3$
dimensions this metric corresponds to the static BTZ black hole \cite{BTZ},
while in $d=4$ it solves the field equations of conformal gravity provided the
base manifold is of constant curvature \cite{Riegert}. This is also so for
conformal gravity in even dimensions \cite{OS}. In odd dimensions, for an
arbitrary base manifold $\Sigma_{d-2}$, the black hole (\ref{bh}) provides a
solution for a special case of Lovelock gravity \cite{DOT2}. This case is such
that the coefficients are fixed so as the theory admits a unique maximally
symmetric AdS vacuum \cite{BHScan} and the Lagrangian can be expressed as a
Chern-Simons form \cite{Chams}. The case of $\Sigma_{d-2}$ of constant
curvature was previously analyzed in \cite{Cai-Soh}, \cite{ATZ} and for
spherical symmetry Eq. (\ref{bh}) reduces to the solution found in
\cite{Dimensionally}. In five dimensions the metric (\ref{bh}) still solves
the field equations even in the presence of a nontrivial fully antisymmetric
torsion \cite{Camello-Fabrizio-Troncoso}. The black hole (\ref{bh}) also
provides a solution for the Lovelock theory in $d=8$ compactified to five
dimensions \cite{Camellazo-Fabriziano}.

\section{Free massive scalar field}

Here it is shown that the Klein-Gordon equation%
\begin{equation}
\left(  \square-m^{2}\right)  \phi=0\ , \label{kg}%
\end{equation}
admits an analytic solution when it is solved on the background metric given
by (\ref{bh}), which allows to find an exact expression for the quasinormal
modes. This can be seen as follows:

The metric (\ref{bh}) can be expressed as%
\begin{equation}
ds^{2}=-\left(  \frac{\hat{r}^{2}}{l^{2}}-1\right)  d\hat{t}^{2}+\frac
{d\hat{r}^{2}}{\frac{\hat{r}^{2}}{l^{2}}-1}+\hat{r}^{2}d\hat{\Sigma}_{d-2}%
^{2}\ ,\label{mlbh}%
\end{equation}
where the time and radial coordinates have been rescaled as $\hat{r}=\frac
{l}{r_{+}}r,$ $\hat{t}=\frac{r_{+}}{l}t$, and the new base manifold
$\hat{\Sigma}_{d-2}$ is related to $\Sigma_{d-2}$ by means of a global
conformal transformation given by $d\hat{\Sigma}_{d-2}^{2}=\frac{r_{+}^{2}%
}{l^{2}}d\Sigma_{d-2}^{2}$. Thus, since the quasinormal modes for a free
scalar field propagating on the metric (\ref{mlbh}) have been already found in
\cite{QNml} requiring the scalar field to be purely ingoing at the horizon and
the vanishing of the energy flux at infinity, the result we are looking for
can be obtained by means of a simple rescaling. Indeed, as for each mode the
scalar field propagating on (\ref{mlbh}) acquires the form
\begin{equation}
\phi=e^{-i\hat{\omega}\hat{t}/l}f\left(  \hat{r}\right)  Y(\hat{\Sigma})\ ,
\end{equation}
where $Y(\hat{\Sigma})$ is an eigenfunction of the Laplace operator on
$\hat{\Sigma}$ with eigenvalue $-\hat{Q}$, i.e. $\nabla_{\hat{\Sigma}}%
^{2}Y(\hat{\Sigma})=-\hat{Q}Y(\hat{\Sigma})$, the scalar field on the black
hole (\ref{bh}), is given by%
\begin{equation}
\phi=e^{-i\omega t/l}f\left(  r\right)  Y(\Sigma)\ ,\label{phi-separation}%
\end{equation}
with%
\begin{align}
\omega &  =\frac{r_{+}}{l}\hat{\omega}\text{\ },\label{ores}\\
Q &  =\frac{r_{+}^{2}}{l^{2}}\hat{Q}\ .\label{Qres}%
\end{align}

Hence, as $\hat{\omega}$ was already found in \cite{QNml}, the quasinormal
frequencies of the black hole (\ref{bh}) turn out to be given by%
\begin{equation}
\omega=-\sqrt{Q-\left(  \frac{d-3}{2}\right)  ^{2}\frac{r_{+}^{2}}{l^{2}}%
}-i\frac{r_{+}}{l}\left(  2n+1+\sqrt{\left(  \frac{d-1}{2}\right)  ^{2}%
+m^{2}l^{2}}\right)  \text{ ,} \label{omega1}%
\end{equation}
where $Q$ is the eigenvalue of the Laplace operator on $\Sigma_{d-2}$, and
$n=0,1,2,...$\ .

The radial function can be expressed in terms of $z=1-r_{+}^{2}/r^{2}$, and it reads%

\begin{equation}
f\left(  z\right)  =z^{\alpha}\left(  1-z\right)  ^{\beta}F\left(
a,b,c,z\right)  \ , \label{fmin}%
\end{equation}
where%
\begin{align}
\alpha &  =-\frac{il\omega}{2r_{+}}\ ,\\
\beta &  =\beta_{\pm}=\frac{d-1}{4}\pm\frac{1}{2}\sqrt{\left(  \frac{d-1}%
{2}\right)  ^{2}+m^{2}l^{2}}\ ,
\end{align}
and $F$ is the hypergeometric function with parameters defined by%
\begin{align}
a  &  =-\left(  \frac{d-3}{4}\right)  +\alpha+\beta_{\pm}+\frac{i}{2}%
\sqrt{\frac{l^{2}}{r_{+}^{2}}Q-\left(  \frac{d-3}{2}\right)  ^{2}}\ ,\\
b  &  =-\left(  \frac{d-3}{4}\right)  +\alpha+\beta_{\pm}-\frac{i}{2}%
\sqrt{\frac{l^{2}}{r_{+}^{2}}Q-\left(  \frac{d-3}{2}\right)  ^{2}}\ ,\\
c  &  =1+2\alpha\ .
\end{align}

Note that, as it occurs for AdS spacetime, stability is guaranteed provided
the Breitenlohner-Freedman bound is fulfilled \cite{BFT}%
\begin{equation}
m^{2}l^{2}\geq-\left(  \frac{d-1}{2}\right)  ^{2}\ .
\end{equation}

From Eq. (\ref{omega1}), one can see that ringing modes exist provided the
following bound on the eigenvalue of the Laplace operator on $\Sigma_{d-2}$ is
fulfilled%
\begin{equation}
Q>\left(  \frac{d-3}{2}\right)  ^{2}\frac{r_{+}^{2}}{l^{2}}\ , \label{boundQ}%
\end{equation}
otherwise the excitations are purely damped.

In the case of spherical symmetry, i.e. for $\Sigma_{d-2}=S^{d-2}$, since
$Q=L\left(  L+d-3\right)  $ the equation (\ref{boundQ}) provides a bound for
the angular momentum $L$ of the perturbation. It is natural then to expect
that this bound should be related with the impact parameter that a geodesic
has to possess in order to avoid being swallowed by the black hole.

Note also that, when (\ref{boundQ}) is fulfilled, the damping time scale is
independent of $Q$, unlike what occurs for the Schwarzschild-AdS black hole,
for which the damping time scale increases with the angular momentum of the
mode \cite{Horowitz-Hubeny}.

Using the results found in this section, it has been recently argued that the
mass and area spectrum of these black holes have a strong dependence on the
base manifold, and they are not evenly spaced \cite{PJL}.

\section{Nonminimal coupling}

Let us consider the following massive scalar field nonminimally coupled with
the scalar curvature%

\begin{equation}
\left(  \square-m^{2}+\xi R\right)  \phi=0\ ,\label{KGnonmin}%
\end{equation}
where the conformal coupling is recovered for $\xi=-\frac{1}{4}\frac{d-2}%
{d-1}$. Here $R$ is the Ricci scalar of the background metric, which for the
black hole (\ref{bh}) is given by%
\begin{equation}
R=-\frac{d\left(  d-1\right)  }{l^{2}}+\frac{l^{2}R_{\Sigma}+\left(
d-2\right)  \left(  d-3\right)  r_{+}^{2}}{l^{2}r^{2}}:=A+\frac{B}{r^{2}%
}\ ,\label{RS}%
\end{equation}
where $R_{\Sigma}$ is the Ricci scalar of the base manifold $\Sigma_{d-2}$,
which hereafter is assumed to be constant in order to ensure the separability
of equation (\ref{KGnonmin}). Unlike the case of case of AdS spacetime, the
Ricci scalar of the black hole (\ref{RS}) is not constant, and hence the
nonminimal coupling contributes now to the field equation with more than just
a shift in the mass. Nevertheless, remarkably, the effect of the nonminimal
coupling amounts to a shift in $Q$, as compared with the previous case. This
can be seen as follows:

Using separation of variables as in Eq. (\ref{phi-separation}), the equation
for the radial function reads%
\begin{equation}
\frac{d^{2}f}{dr^{2}}+\left[  \frac{d-2}{r}+\frac{2r}{r^{2}-r_{+}^{2}}\right]
\frac{df}{dr}+\left[  \frac{\omega^{2}l^{4}}{\left(  r^{2}-r_{+}^{2}\right)
^{2}}-\frac{l^{2}\left(  Q-\xi B\right)  }{\left(  r^{2}-r_{+}^{2}\right)
r^{2}}-\frac{l^{2}\left(  m^{2}-\xi A\right)  }{r^{2}-r_{+}^{2}}\right]
f=0\ , \label{fkgnm}%
\end{equation}

Note that one obtains the same equation as in the case of minimal coupling
($\xi=0$), which has already been solved in the previous section, but with an
effective mass and "angular momentum" given by%
\begin{align}
Q_{eff}  &  =Q-\left(  R_{\Sigma}+\left(  d-2\right)  \left(  d-3\right)
\frac{r_{+}^{2}}{l^{2}}\right)  \xi\ ,\\
m_{eff}^{2}  &  =m^{2}+\frac{d\left(  d-1\right)  }{l^{2}}\xi\ .
\end{align}

Therefore, the solution of Eq. (\ref{fkgnm}) can be written as in
(\ref{phi-separation}) with (\ref{fmin}), replacing $Q$ and $m^{2}$ by
$Q_{eff}$ and $m_{eff}^{2}$, respectively. The quasinormal frequencies are
then given by (\ref{omega1}) with the same replacements. The presence of a
nonminimal coupling changes the boundary condition on the vanishing of the
energy flux at infinity, such that it can be compatible with scalar fields
possessing slow fall-off. If the mass and the coupling constant $\xi$ satisfy
the relation%
\begin{equation}
\xi+\beta+4\xi\beta=0\ ,
\end{equation}
with%
\begin{equation}
\beta:=\frac{d-1}{4}\pm\frac{1}{2}\sqrt{\left(  \frac{d-1}{2}\right)
^{2}+m_{eff}^{2}l^{2}}\ ,
\end{equation}
then for the range of effective masses given by%
\begin{equation}
-\left(  \frac{d-1}{2}\right)  ^{2}<m_{eff}^{2}l^{2}<1-\left(  \frac{d-1}%
{2}\right)  ^{2}\ ,
\end{equation}
there is a second set of modes for which the frequencies can be obtained from
Ref. \cite{QNml}. Thus, using the corresponding scalings and shifts explained
in this section, the second set of quasinormal frequencies turns out to be%
\begin{equation}
\omega=-\sqrt{Q_{eff}-\left(  \frac{d-3}{2}\right)  ^{2}\frac{r_{+}^{2}}%
{l^{2}}}-i\frac{r_{+}}{l}\left(  2n+1-\sqrt{\left(  \frac{d-1}{2}\right)
^{2}+m_{eff}^{2}l^{2}}\right)  \ .
\end{equation}

\section{Electromagnetic field}

The quasinormal modes for an electromagnetic perturbation that propagates on
the black hole (\ref{bh}) can be obtained following the same strategy as the
one for the scalar field. Indeed, the quasinormal modes for the Maxwell field
propagating on the massless topological black hole (\ref{mlbh}) have been
found by L\'{o}pez-Ortega in Ref. \cite{LO}, who showed that the problem can
be reduced to the case of the scalar field solved in Ref. \cite{QNml}, since
scalar and vector modes of the electromagnetic perturbation are equivalent to
scalar field perturbations but with different precise masses. Therefore by
virtue of Eqs. (\ref{ores}) and (\ref{Qres}), the quasinormal frequencies for
the scalar and vector modes of the electromagnetic perturbation on the black
hole (\ref{bh}) are respectively given by%
\begin{align}
d &  =4:\omega_{s}=-\sqrt{Q_{s}-\frac{1}{4}\frac{r_{+}^{2}}{l^{2}}}%
-2i\frac{r_{+}}{l}\left(  n+\frac{3}{4}\right)  \ ,\\
d &  \geq5:\omega_{s}=-\sqrt{Q_{s}-\left(  \frac{d-3}{2}\right)  ^{2}%
\frac{r_{+}^{2}}{l^{2}}}-2i\frac{r_{+}}{l}\left(  n+\frac{3}{4}\right)  \ ,
\end{align}
and%
\begin{equation}
d\geq4:\omega_{v}=-\sqrt{Q_{v}-\left[  1+\left(  \frac{d-3}{2}\right)
^{2}\right]  \frac{r_{+}^{2}}{l^{2}}}-2i\frac{r_{+}}{l}\left(  n+\frac{d-1}%
{4}\right)  \ ,
\end{equation}
where $Q_{s}$ and $Q_{v}$ are the eigenvalues of the Laplacian on the base
manifold $\Sigma_{d-2}$ for scalar and vector harmonics, respectively.

\bigskip

\textit{Acknowledgments.} We thank S. Dain, G. Dotti, R. Gleiser and the
organizers of the conference: "50 Years of FaMAF and Workshop on Global
Problems in Relativity", hosted during November 2006 at FaMAF, Universidad
Nacional de C\'{o}rdoba, C\'{o}rdoba, Argentina, for the opportunity of
presenting part of this work as a plenary talk. This research is partially
funded by Fondecyt grants N%
${{}^o}$
1061291, 1071125, 1085322, 1095098, 11090281, and to the Conicyt grant
"Southern Theoretical Physics Laboratory" ACT-91. The Centro de Estudios
Cient\'{\i}ficos (CECS) is funded by the Chilean Government through the
Millennium Science Initiative and the Centers of Excellence Base Financing
Program of Conicyt. CECS is also supported by a group of private companies
which at present includes Antofagasta Minerals, Arauco, Empresas CMPC, Indura,
Naviera Ultragas and Telef\'{o}nica del Sur. CIN is funded by Conicyt and the
Gobierno Regional de Los R\'{\i}os.

\end{document}